\documentclass[aps,
prl,
twocolumn,
superscriptaddress,
amsmath
]{revtex4-2}

\usepackage[utf8]{inputenc}
\usepackage{graphicx}
\usepackage{placeins}
\usepackage{float}
\usepackage{subfigure}
\usepackage{longtable}
\usepackage{amssymb}
\usepackage{filecontents}
\usepackage{natbib}
\usepackage{xcolor}
\usepackage{hyperref}
\usepackage{amsmath}
\usepackage{siunitx}
\usepackage{xcolor}
\usepackage{csquotes}
\usepackage{nicefrac}
\DeclareUnicodeCharacter{2212}{\textendash}
\hypersetup{
    colorlinks=true,
    linkcolor=blue,
    filecolor=magenta,      
    urlcolor=blue,
    citecolor=blue,
}

\begin{document}

\title{Charge and spin order dichotomy in NdNiO$_2$ driven by SrTiO$_3$ capping layer}

\author{G. Krieger}\thanks{These authors contributed equally}
\affiliation{Université de Strasbourg, CNRS, IPCMS UMR 7504, F-67034 Strasbourg, France}
\author{L. Martinelli}\thanks{These authors contributed equally}
\affiliation{Dipartimento di Fisica, Politecnico di Milano, Piazza Leonardo da Vinci 32, I-20133 Milano, Italy}

\author{S. Zeng}
\affiliation{Department of Physics, Faculty of Science, National University of Singapore, 117551 Singapore, Singapore}

\author{L. E. Chow}
\affiliation{Department of Physics, Faculty of Science, National University of Singapore, 117551 Singapore, Singapore}

\author{K. Kummer}
\affiliation{ESRF, The European Synchrotron, 71 Avenue des Martyrs, F-38043 Grenoble, France}

\author{R. Arpaia}
\affiliation{Quantum Device Physics Laboratory, Department of Microtechnology and Nanoscience, Chalmers University of Technology, SE-41296 G$\ddot{o}$teborg, Sweden}

\author{M. Moretti Sala}
\affiliation{Dipartimento di Fisica, Politecnico di Milano, Piazza Leonardo da Vinci 32, I-20133 Milano, Italy}

\author{N.B. Brookes}
\affiliation{ESRF, The European Synchrotron, 71 Avenue des Martyrs, F-38043 Grenoble, France}

\author{A. Ariando}
\affiliation{Department of Physics, Faculty of Science, National University of Singapore, 117551 Singapore, Singapore}

\author{N. Viart}
\affiliation{Université de Strasbourg, CNRS, IPCMS UMR 7504, F-67034 Strasbourg, France}

\author{M. Salluzzo}\email[]{marco.salluzzo@spin.cnr.it}
\affiliation{CNR-SPIN Complesso di Monte S. Angelo, via Cinthia - I-80126 Napoli, Italy}
\author{G. Ghiringhelli}\email[]{giacomo.ghiringhelli@polimi.it}
\affiliation{Dipartimento di Fisica, Politecnico di Milano, Piazza Leonardo da Vinci 32, I-20133 Milano, Italy}
\affiliation{CNR-SPIN, Dipartimento di Fisica, Politecnico di Milano, Piazza Leonardo da Vinci 32, I-20133 Milano, Italy}
\author{D. Preziosi}\email[]{daniele.preziosi@ipcms.unistra.fr}
\affiliation{Université de Strasbourg, CNRS, IPCMS UMR 7504, F-67034 Strasbourg, France}

\begin{abstract}
  Superconductivity in infinite-layer nickelates holds exciting analogies with that of cuprates, with similar structures and $3d$-electron count.
  Using resonant inelastic x-ray scattering (RIXS) we studied electronic and magnetic excitations and charge density correlations in Nd$_{1-x}$Sr$_{x}$NiO$_2$ thin films with and without an SrTiO$_3$ capping layer. We observe dispersing magnons only in the capped samples, progressively dampened at higher doping. In addition, we detect an elastic resonant scattering peak in the uncapped $x=0$ compound at wave vector $(\nicefrac{1}{3},0)$, remindful of the charge order signal in hole doped cuprates. The peak weakens at $x=0.05$ and disappears in the superconducting $x=0.20$ film. The uncapped samples also present a higher degree of Ni$3d$-Nd$5d$ hybridization and a smaller anisotropy of the Ni$3d$ occupation with respect to the capped samples. The role of the capping on the possible hydrogen incorporation or on other mechanisms responsible for the electronic reconstruction far from the interface remains to be understood.  
\end{abstract}

\maketitle

The prediction of high $T_{c}$ superconductivity in nickelate-based heterostructures, in particular LaNiO$_{3}$/LaAlO$_{3}$ superlattices \cite{Anisimov1999,Chaloupka2008}, has stimulated an important experimental activity aiming at mimicking the electronic structure of CuO$_{2}$ planes of cuprates. There, superconductivity is linked to an inherent 2D electronic structure with $(x^2-y^2)$ symmetry at all doping levels \cite{Oles2019} and to the spin $\nicefrac{1}{2}$ atomic momentum of Cu-$3d^9$ sites. 
Special efforts have thus been deployed to stabilize a Ni-$3d^9$ configuration with reduced dimensionality. A partial splitting of the Ni-$e_{g}$ states was obtained in strained LaNiO$_{3}$ thin films \cite{Chakhalian2011}, LaTiO$_{3}$/LaNiO$_{3}$/LaAlO$_{3}$ heterostructures \cite{Disa2015}, LaNiO$_{3}$/LaAlO$_{3}$ multilayers \cite{Benckiser2011} and LaTiO$_{3-\delta}$/LaNiO$_{3}$ bilayers \cite{Cao2016}, but no superconductivity was reported. Only recently a zero-resistance state was found below 15K in Nd$_{0.8}$Sr$_{0.2}$NiO$_{2}$ thin films deposited on the (001) surface of SrTiO$_3$ (STO) \cite{Li2019}. The main step to achieve this result was an oxygen de-intercalation of the pristine perovskite Nd$_{1-x}$Sr$_{x}$NiO$_3$ phase via a topotactic reduction by using a CaH$_2$ powder as reagent. When properly realized, this soft chemistry process leads to an infinite-layer structure with Ni-$3d^{(9-x)}$ ions in a square planar lattice, in all similar to that of hole doped cuprates.

Soon after this discovery, electron energy-loss spectroscopy \cite{Goodgee2007683118}, x-ray absorption spectroscopy (XAS) and resonant inelastic x-ray scattering (RIXS) experiments on undoped (LaNiO$_2$, NdNiO$_2$) \cite{Hepting2020} and doped (Nd$_{1-x}$Sr$_{x}$NiO$_2$) samples \cite{Rossi2020} showed that the O-$2p$ and Ni-$3d$ bands are more separated in energy in the nickelate infinite-layer ($\sim$4 eV) than in layered cuprates ($\sim$2 eV) \cite{BotanaNorman2020}. Although still debated \cite{Zhang2020, Millis2020, Sawatzky2020}, these results would place nickelate parent compounds in the Mott-Hubbard region of the  Zaanen–Sawatzky–Allen classification scheme, implying that injecting holes leads towards a Ni-$3d^8$ electronic configuration. This is at odds with cuprates, classified as charge-transfer insulators, where doping holes go to the oxygen sites and form the so called Zhang-Rice singlets, Cu-$3d^9\underline{L}$  \cite{Pellegrin1993,Brookes2001}.
Moreover, in infinite-layer nickelates a sizable contribution of the Nd-$5d$ bands at the Fermi level leads to an important hybridization of Nd-$5d$ and Ni-$3d$ states \cite{BotanaNorman2020,Hepting2020}, whereas in cuprates the inter-layer cations do not contribute to the electronic states relevant to transport. 
Besides these differences, the analogy with cuprates was recently strengthened by the observation of magnons in NdNiO$_2$ \cite{Lu213}, dispersing similarly to those of cuprates, but on a smaller energy range. 
Interestingly, magnons were found so far only in nickelates capped with epitaxial STO before the topotactic reduction, whereas the capping is not necessary to get superconductivity \cite{Ariando2020}. An important feature of all cuprates is the charge density instability of hole doped CuO$_2$ planes \cite{emery1993frustrated, castellani1995singular, castellani1996non}. 
\begin{figure*}[tt]
\centering
\includegraphics[width=1.00\textwidth]{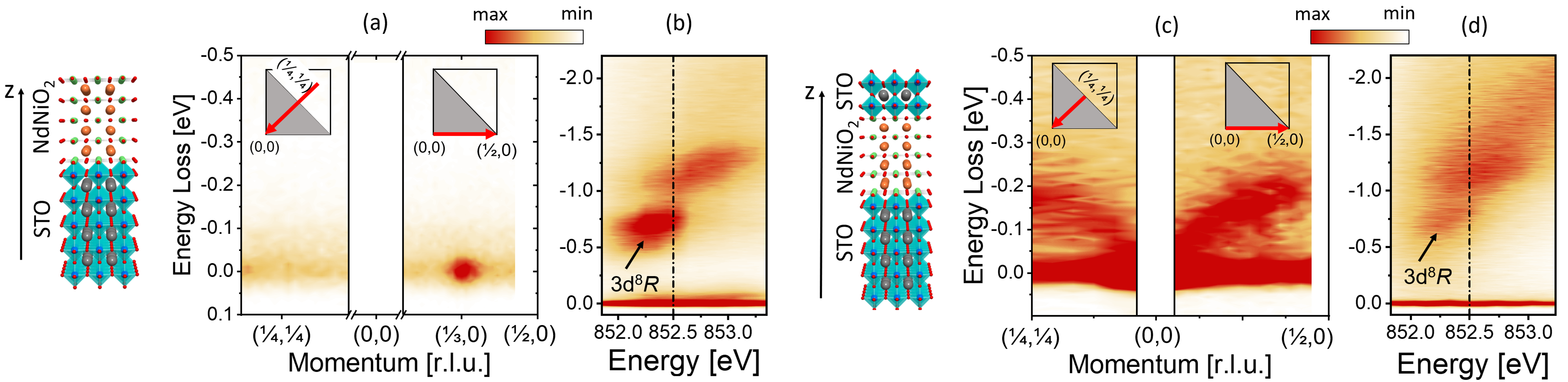}
\caption{\label{fig1} Summary of the RIXS results for uncapped and STO capped NSNO(0) films. (a,c): energy-loss/in-plane-momentum scattering intensity maps along the high symmetry directions indicated in the insets, excited at incident photon energy $\sim$852.5 eV (Ni$^{1+}$ XAS peak) using $\pi$ polarization. (b,d): energy-loss/excitation-energy maps across the Ni L$_3$ edge, at 10$^\circ$ grazing incidence. Lateral panels show sketches of the structure of both (left) uncapped and (right) capped samples.
}
\end{figure*}
It emerges in the form of charge density waves (CDW) in selected regions of the temperature/doping ($T/p$) phase diagram \cite{Ghiringhelli821, achkar2012distinct, CD2016, wahlberg2021restored}, and of charge density fluctuations (CDF) over a more extended set of $T$ and $p$ values \cite{Arpaia906,seibold2021strange,Riccardo}. To our knowledge, CDW and/or CDF have not been reported so far in infinite-layer nickelate thin films.  

In this work, we use RIXS at the Ni L$_3$-edge to study the spin and charge correlations in the NiO$_2$ planes of infinite-layer nickelates. We compare data of Nd$_{1-x}$Sr$_{x}$NiO$_2$ thin films (NSNO(x)) at three doping levels ($x$=0.00, 0.05, 0.20), with and without the STO capping layer. We confirm the presence of spin-spin correlation by detecting dispersing magnons in capped samples, and we observe a resonant zero-energy-loss scattering signal peaking at $(\nicefrac{1}{3},0)$ [r.l.u.] in uncapped samples. We also find other relevant differences related to the presence of the capping layer in the RIXS and XAS spectra, that we will thoroughly address in the following.

NSNO(x) samples were obtained via a topotactic reduction process of 10 nm thick perovskite precursor films grown by pulsed laser deposition. STO (3 unit cells) capped  and uncapped NSNO(0) and NSNO(0.05) samples were prepared at IPCMS in Strasbourg following the procedure described in Ref. \cite{Preziosi2017} and Supplemental Materials \cite{SI}, whereas superconducting uncapped NSNO(0.2) samples were made at NUS in Singapore \cite{Ariando2020}. X-ray diffraction data demonstrate the formation of the infinite-layer structure, with similar $c$-axis values for capped and uncapped samples at each doping \cite{SI}. The temperature dependence of the resistivity is in agreement with literature \cite{Li2019}, with a metallic behavior and low temperature upturn in NSNO(0) and NSNO(0.05), and zero resistance transition in NSNO(0.2). However, depending on the details of the topotactic procedure, some uncapped NSNO(0) can also exhibit a semiconducting temperature dependence in the whole temperature range (4-300 K) \cite{SI}, similarly to bulk NdNiO$_2$ \cite{Li2020}, and to early reported data on LaNiO$_2$ thin films \cite{Hepting2020}. The XAS and RIXS measurements were performed at the beam line ID32 of the European Synchrotron -- ESRF (Grenoble, France) with the ERIXS spectrometer \cite{Brookes2018}. RIXS spectra were acquired with a combined (beam line + spectrometer) energy resolution of $\sim42$~meV at the Ni L$_3$-edge ($\sim 853$~eV). The sample surface was perpendicular to the scattering plane and the linear polarization of the incident photons was either parallel ($\pi$) or perpendicular ($\sigma$) to it. The projection $\mathbf{q_\parallel}$ of the transferred momentum $\mathbf{q}=\mathbf{k'}-\mathbf{k}$ in the NiO$_2$ planes was changed by rotating the sample around an axis perpendicular to the scattering plane at fixed scattering angle $2\theta=149.5^\circ$. The high symmetry (10) ($\Gamma-$X) and (11) ($\Gamma-$Y) directions in the 2D Brillouin zone (BZ) were explored.

Figures ~\ref{fig1}(a,c) present the dispersion of low energy excitations in uncapped and capped NSNO(0) samples, respectively. Insets show sketches of (left) uncapped and (right) capped NSNO(0) samples. The excitation energy of $\sim$852.5~eV corresponds to the main Ni$^{1+}$ XAS resonance. Capped NSNO(0) shows strong magnetic excitations dispersing up to $\sim 200$~meV at the BZ boundary, in agreement with recent results \cite{Lu213}. On the contrary, in the uncapped NSNO(0) there are no features attributable to dispersing magnons and we find a strongly modulated intensity of the quasi-elastic region along the (10) direction. This feature is strongly resonant, as it disappears when the incident photon energy is detuned by few hundreds meV from the XAS peak \cite{SI}. The two samples differ also in the intermediate energy-loss region presented in Figs. ~\ref{fig1}(b,d). We can notice that the excitation at around -0.6~eV, commonly associated to the 3d$^{8}R$ final state where an electron is transferred from Ni-$3d$ to Nd-$5d$, is much more intense in uncapped NSNO(0), while the fluorescence-like tail at higher energy is more pronounced for the capped one. 

\begin{figure}[t]
\centering
\includegraphics[width=0.39\textwidth]{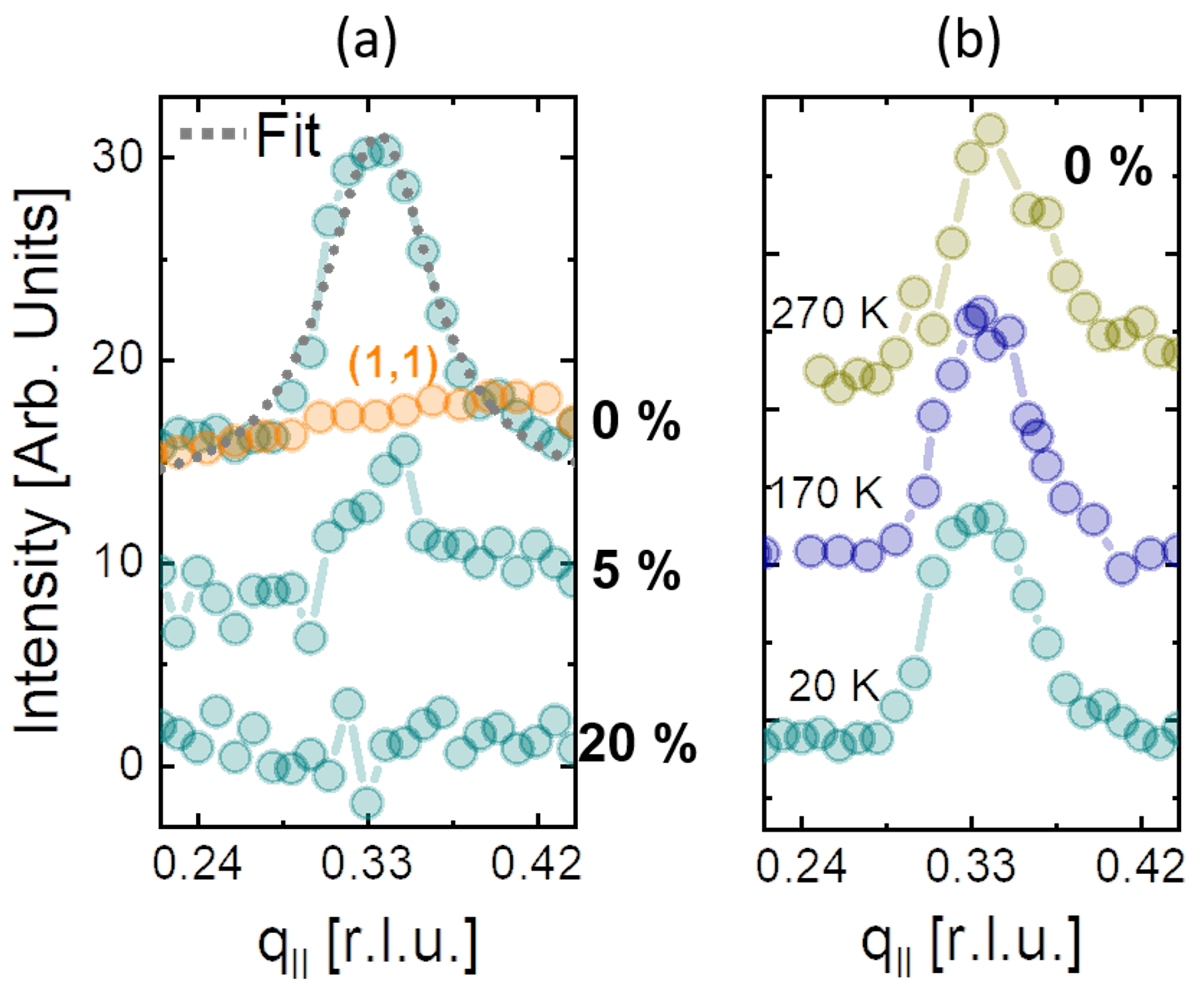}
\caption{\label{fig2} (a) Quasi-elastic scattering intensity vs. the in-plane transferred momentum along the (10) direction for different doping in uncapped NSNO(x). Data acquired at 20~K and grazing incidence with $\pi$ polarization. Corresponding data along the (11) direction (in orange) is presented for comparison as well. (b) Temperature dependence of the quasi-elastic peak area along (10) for uncapped NSNO(0).}
\end{figure} 

To gain insights on the nature of the resonant quasi-elastic peak, we report in Fig.~\ref{fig2} its integrated intensity vs $q_{\parallel}$ along (10), obtained by fitting the RIXS spectra in the [0.1, -0.5]~eV energy range with one Gaussian and one asymmetric Lorentzian at zero and $\sim$60 meV energy loss, respectively; we assign the latter to an optical phonon due to its small dispersion \cite{SI}. For NSNO(0) the $q_{\parallel}$ scan can be fitted by a single Lorentzian centered at $H=0.33$ reciprocal lattice units (r.l.u.), with a full width at half maximum (FWHM) of about $\sim0.06$ r.l.u., corresponding to a real space periodicity of $\sim 3a = 11.7$~\AA~ and to a correlation length $\frac{1}{\pi \cdot FWHM}$ of $ \sim 5.3 a = 20.7$ \AA. Importantly, no modulation of the quasi-elastic intensity is observed along the (11) direction and as a function of the perpendicular momentum, suggesting a two-dimensional nature of the ordering \cite{SI}, reminiscent of the CDW in cuprates. Upon doping, the signal quickly weakens by approximately a factor 3 for $x=0.05$ and disappears in the $x=0.20$ sample. Moreover, the peak is insensitive to temperature in the 20-270 K range. The resonant nature and the peculiar $q$-dependence suggest close analogies with the charge order (CO) phenomenon observed in all hole-doped cuprates \cite{Ghiringhelli821,CD2016,Arpaia906}. However the doping dependence is markedly different, because in cuprates charge ordering is absent below $p = 0.07$ and is maximum at $p \sim 0.11-0.13$, whereas in nickelates it is maximum at zero doping. Moreover the peak position at $q \sim (\nicefrac{1}{3},0)$ is compatible with a commensurate modulation with period $3a$, which has not been observed in cuprates. Also the temperature dependence is different from that commonly found in cuprates; in fact a narrow $T$-independent peak has been observed only in overdoped (Bi,Pb)$_{2.12}$Sr$_{1.88}$CuO$_{6+\delta}$ (Bi2201) family \cite{Peng2018}. This partial resemblance with the phenomenology of the CO in cuprates, tells us that the actual origin of this modulation of the quasi-elastic resonant scattering signal in infinite-layer nickelates is not obvious, as discussed below.

\begin{figure}[ht]
\centering
\includegraphics[width=0.48\textwidth]{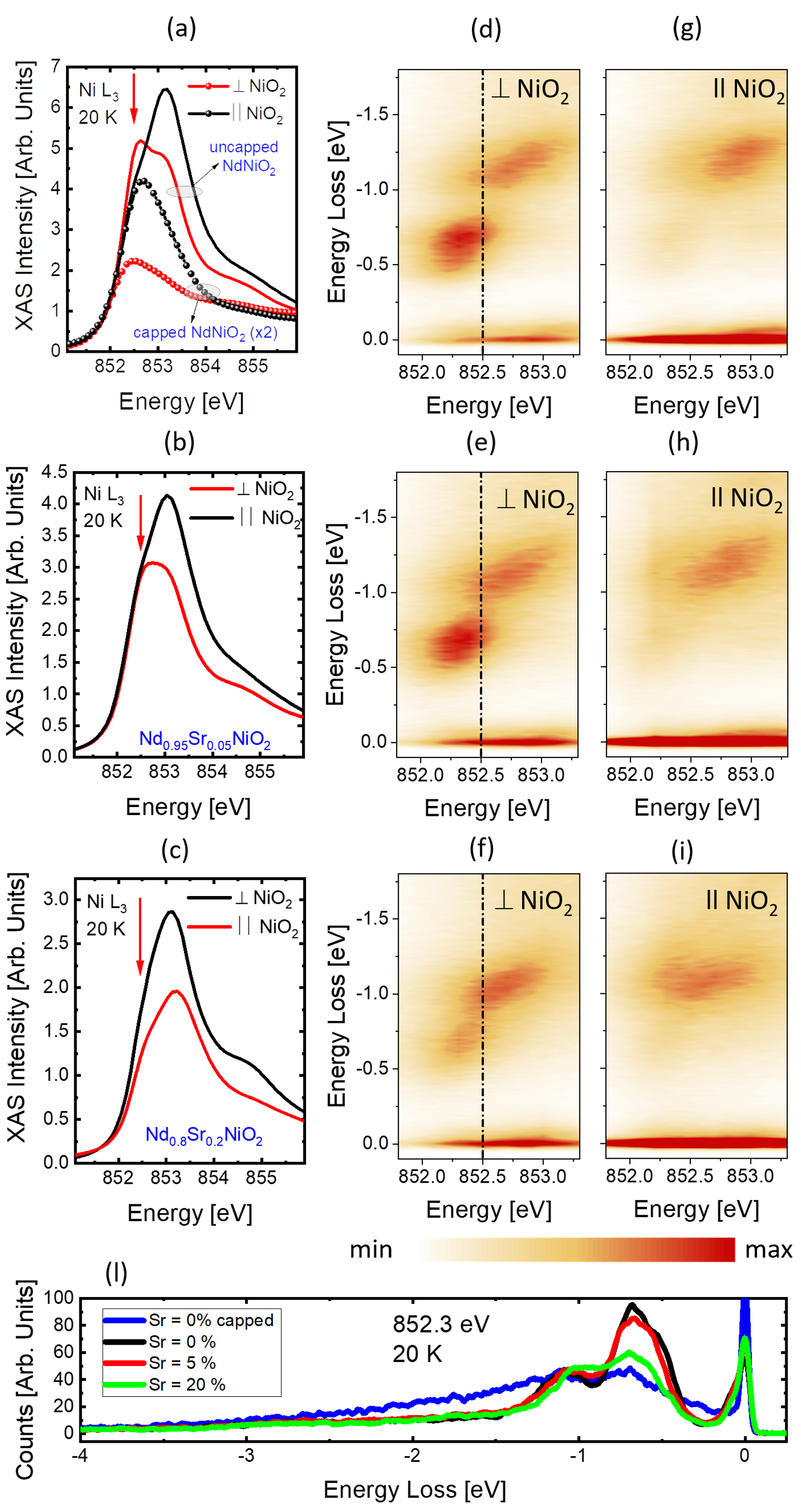}
\caption{\label{fig3} Anisotropy of the XAS and RIXS signal in uncapped NSNO(x) films with $x=0.00, 0.05, 0.20$. Spectra were acquired with linear polarization parallel and at 80$^\circ$ to the NiO$_{2}$ planes, labelled for brevity $\parallel$NiO$_2$ and $\perp$NiO$_2$ respectively. For comparison, in (a) we show also the corresponding XAS spectra of capped NSNO(0). The RIXS spectra of panel (l), excited at 852.5~eV in $\perp$NiO$_2$ geometry, show that the 3d$^{8}R$ -0.6 eV feature decreases with doping in the uncapped samples and is weakest in the capped one. 
}
\end{figure}

We use the incident photon polarization dependence of XAS and RIXS spectra to devise the orbital character of the electronic states. Fig.~\ref{fig3}(a) shows a comparison of the XAS spectra of capped and uncapped NSNO(0). The spectra were measured at 10$^\circ$ grazing incidence, with linear polarization forming an angle of 0$^\circ$ or 80$^\circ$ to the NiO$_{2}$ planes, and labelled for brevity $\parallel$NiO$_2$ and $\perp$NiO$_2$, respectively. While for capped samples the main XAS peak is at an energy ($\sim$ 852.5 eV) corresponding to the nominal Ni$^{1+}$ ground state, in the uncapped samples the maximum of the XAS intensity is at higher energy ($\sim$ 853 eV), which can be assigned to the nominal Ni$^{2+}$ configuration upon comparison with reference spectra of NiO.
In the uncapped samples the Ni$^{1+}$ feature shows up as a shoulder on the low-energy side of the main peak, it is relatively more prominent in the $\perp$NiO$_{2}$ configuration, and its relative weight with respect to the Ni$^{2+}$ peak decreases with doping, as one should expect. The comparison of the XAS spectra in the two geometries tells that the in-plane $3d$ states are less occupied than out-of-plane ones and that this orbital anisotropy is stronger in the capped than in the uncapped samples, in agreement with literature \cite{Rossi2020,Lu213}. The differences in the electronic structure revealed by the XAS spectra echo in the RIXS maps of Figs.~\ref{fig3}(d)-(i). In the uncapped films the 3d$^{8}R$ peak around -0.6 eV energy loss is indeed very strong only in the $\perp$NiO$_2$ geometries and resonates at the Ni$^{1+}$ XAS peak. Moreover, it decreases as a function of the Sr-doping similarly to the Ni$^{1+}$ shoulder in the XAS spectra, which suggests a direct correlation between the large Ni-Nd hybridization and the substantial contribution to the ground-state of the out-of-plane $3d$ orbitals. We can also note that the 3d$^{8}R$ feature measured in $\perp$NiO$_{2}$ geometry is much smaller in the capped than in the uncapped NSNO(0) sample. 

The experimental results poses several questions about the relationship between the electronic properties revealed by XAS and RIXS and the quasi-elastic scattering intensity modulation. In particular, it is unclear whether it should be attributed to  a charge density wave phenomenon or to a superstructure in the lattice, and why it is suppressed in capped samples which, on the other hand, show dispersing magnon-excitations. For simplicity in the following discussion we will refer to the quasi-elastic scattering intensity modulation as charge order (CO) peak. From our study it emerges an intriguing correlation between a strong out-of-plane Ni-Nd hybridization and the observation of CO in undoped and slightly Sr-doped infinite-layer nickelate thin films, which is difficult to understand at this stage. According to our data, the resonating character of the CO peak implies that it has to do with a spatial charge modulation within the NiO$_2$ planes. In undoped NSNO(0) this implies an intrinsic hole doping of the planes, which can be explained in terms of an electron transfer from Ni towards Nd states, hence a self-doping the system; recent DFT calculations \cite{Millis2020} estimated that this effect could account for as much as 0.1 holes/Ni in bulk NdNiO$_2$. Interestingly, we also observed the CO peak at the Nd M$_5$ edge \cite{SI} which, although probing Nd-$4f$ states, indirectly reflects a charge modulation of the Nd-$5d$ states. Thus, the picture that emerges is a correlation between the strong Ni-Nd out-of-plane hybridization and CO, which both compete with magnetic excitation and, ultimately, with superconductivity. Thus, the physics of cuprates is effectively re-established in infinite-layer nickelates only if the Nd-$5d$ states are sufficiently far in energy from the Ni-$3d_{z^2}$ states.

On the other hand, the CO is observed irrespective of the insulating and/or bad-metal transport properties of the samples, although it is more intense for the insulating ones \cite{SI}. It is then essential to ask ourselves to which extent the presence/absence of the STO capping layer, by triggering possible structural and/or chemical modifications of the surface/interface states, results in very unique electronic properties. From the structural analysis of the two sample families, we do not find relevant differences in the average distance between the NiO$_{2}$ planes. However, differences might be induced by the kinetics of the topotactic reaction, influenced by the presence of the capping layer. In particular, the topotactic de-intercalation of oxygen ions could induce the inclusion of hydrogen ions released from the CaH$_2$ powder. According to density functional theory (DFT) calculations \cite{TopoTheory2020}, the NdNiO$_2$H phase, is favourable in NdNiO$_2$, while unfavourable in Sr-doped NdNiO$_2$. The two phases differ substantially in their electronic properties. In particular, the Nd-$5d$ electron pocket at the Fermi level disappears in the case of the NdNiO$_2$H phase with a predominant Ni$^{2+}$ valence state, and almost equal occupancy of $3d_{z^2}$ and $3d_{x^2-y^2}$ orbitals. Some of these electronic characteristics are, indeed, observed in uncapped NSNO(x). Therefore, we cannot discard the possibility that H atoms get organized as a superstructure, in itself invisible to x-ray diffraction, but inducing a modulation of the electronic structure that results in a CO peak. Moreover the presence of H atoms might reduce the occupation anisotropy of Ni-$3d$ states by acting as a bridge to the Nd planes. However, the calculations also predict a reduced hybridization between Ni-$3d_{z^2}$ and Nd-$5d$ orbitals in the presence of H atoms, while our uncapped samples show experimentally a very strong 3d$^{8}R$ peak resonating at an absorption energy corresponding to Ni$^{1+}$ and the 3d$^{8}R$ intensity and the presence of CO seem to be correlated.

These results suggest that the stability of the superconducting state in nickelates could be effectively extended by reducing the role of the rare-earth electronic states on the NiO$_2$ planes. An increase of the NiO$_2$ inter-planar distance realized, for example, via uniaxial pressure, chemical substitution of Nd with larger ionic radii rare-earths or by epitaxial in-plane compressive strain, could represent viable routes.

\section{acknowledgments}
This work was funded by the French National Research Agency (ANR) through the ANR MISSION ANR-18-CE-CE24-0008-01. D.~P. has benefited support from the initiative of excellence IDEX-Unistra (ANR-10-IDEX-0002-02) from the French national program Investment for the future. L.~M., M.~M.~S. and G.~G. acknowledge support by the project PRIN2017 “Quantum-2D” ID 2017Z8TS5B of the Ministry for University and Research (MIUR) of Italy. The work at NUS is supported by the Agency for Science, Technology, and Research (A*STAR) under its Advanced Manufacturing and Engineering (AME) Individual Research Grant (IRG) (A1983c0034) and by the Singapore Ministry of Education (MOE) Under its Tier 2 program. R.~A. acknowledges support  by the Swedish Research Council (VR) under the Project 2020-04945. The synchrotron experiments were performed at ESRF synchrotron facility in France under proposal number HC-4166 and HC-4628.
\section{Bibliography}
\bibliography{prl.bib}

\providecommand{\noopsort}[1]{}\providecommand{\singleletter}[1]{#1}%
\begin{thebibliography}{35}%
\makeatletter
\providecommand \@ifxundefined [1]{%
 \@ifx{#1\undefined}
}%
\providecommand \@ifnum [1]{%
 \ifnum #1\expandafter \@firstoftwo
 \else \expandafter \@secondoftwo
 \fi
}%
\providecommand \@ifx [1]{%
 \ifx #1\expandafter \@firstoftwo
 \else \expandafter \@secondoftwo
 \fi
}%
\providecommand \natexlab [1]{#1}%
\providecommand \enquote  [1]{``#1''}%
\providecommand \bibnamefont  [1]{#1}%
\providecommand \bibfnamefont [1]{#1}%
\providecommand \citenamefont [1]{#1}%
\providecommand \href@noop [0]{\@secondoftwo}%
\providecommand \href [0]{\begingroup \@sanitize@url \@href}%
\providecommand \@href[1]{\@@startlink{#1}\@@href}%
\providecommand \@@href[1]{\endgroup#1\@@endlink}%
\providecommand \@sanitize@url [0]{\catcode `\\12\catcode `\$12\catcode
  `\&12\catcode `\#12\catcode `\^12\catcode `\_12\catcode `\%12\relax}%
\providecommand \@@startlink[1]{}%
\providecommand \@@endlink[0]{}%
\providecommand \url  [0]{\begingroup\@sanitize@url \@url }%
\providecommand \@url [1]{\endgroup\@href {#1}{\urlprefix }}%
\providecommand \urlprefix  [0]{URL }%
\providecommand \Eprint [0]{\href }%
\providecommand \doibase [0]{https://doi.org/}%
\providecommand \selectlanguage [0]{\@gobble}%
\providecommand \bibinfo  [0]{\@secondoftwo}%
\providecommand \bibfield  [0]{\@secondoftwo}%
\providecommand \translation [1]{[#1]}%
\providecommand \BibitemOpen [0]{}%
\providecommand \bibitemStop [0]{}%
\providecommand \bibitemNoStop [0]{.\EOS\space}%
\providecommand \EOS [0]{\spacefactor3000\relax}%
\providecommand \BibitemShut  [1]{\csname bibitem#1\endcsname}%
\let\auto@bib@innerbib\@empty
\bibitem [{\citenamefont {Anisimov}\ \emph {et~al.}(1999)\citenamefont
  {Anisimov}, \citenamefont {Bukhvalov},\ and\ \citenamefont
  {Rice}}]{Anisimov1999}%
  \BibitemOpen
  \bibfield  {author} {\bibinfo {author} {\bibfnamefont {V.~I.}\ \bibnamefont
  {Anisimov}}, \bibinfo {author} {\bibfnamefont {D.}~\bibnamefont
  {Bukhvalov}},\ and\ \bibinfo {author} {\bibfnamefont {T.~M.}\ \bibnamefont
  {Rice}},\ }\bibfield  {title} {\bibinfo {title} {{Electronic structure of
  possible nickelate analogs to the cuprates}},\ }\href
  {https://doi.org/10.1103/PhysRevB.59.7901} {\bibfield  {journal} {\bibinfo
  {journal} {Phys. Rev. B}\ }\textbf {\bibinfo {volume} {59}},\ \bibinfo
  {pages} {7901} (\bibinfo {year} {1999})}\BibitemShut {NoStop}%
\bibitem [{\citenamefont {Chaloupka}\ and\ \citenamefont
  {Khaliullin}(2008)}]{Chaloupka2008}%
  \BibitemOpen
  \bibfield  {author} {\bibinfo {author} {\bibfnamefont {J.}~\bibnamefont
  {Chaloupka}}\ and\ \bibinfo {author} {\bibfnamefont {G.}~\bibnamefont
  {Khaliullin}},\ }\bibfield  {title} {\bibinfo {title} {{Orbital order and
  possible superconductivity in LaNiO$_3$/LaMO$_3$ superlattices}},\ }\href
  {https://doi.org/10.1103/PhysRevLett.100.016404} {\bibfield  {journal}
  {\bibinfo  {journal} {Phys. Rev. Lett.}\ }\textbf {\bibinfo {volume} {100}},\
  \bibinfo {pages} {3} (\bibinfo {year} {2008})}\BibitemShut {NoStop}%
\bibitem [{\citenamefont {Ole\'{s}}\ \emph {et~al.}(2019)\citenamefont
  {Ole\'{s}}, \citenamefont {Wohlfeld},\ and\ \citenamefont
  {Khaliullin}}]{Oles2019}%
  \BibitemOpen
  \bibfield  {author} {\bibinfo {author} {\bibfnamefont {A.~M.}\ \bibnamefont
  {Ole\'{s}}}, \bibinfo {author} {\bibfnamefont {K.}~\bibnamefont {Wohlfeld}},\
  and\ \bibinfo {author} {\bibfnamefont {G.}~\bibnamefont {Khaliullin}},\
  }\bibfield  {title} {\bibinfo {title} {{Orbital Symmetry and Orbital
  Excitations in High-Tc Superconductors}},\ }\href
  {https://doi.org/10.3390/condmat4020046} {\bibfield  {journal} {\bibinfo
  {journal} {Condens. Matter}\ }\textbf {\bibinfo {volume} {4}},\ \bibinfo
  {pages} {46} (\bibinfo {year} {2019})}\BibitemShut {NoStop}%
\bibitem [{\citenamefont {Chakhalian}\ \emph {et~al.}(2011)\citenamefont
  {Chakhalian}, \citenamefont {Rondinelli}, \citenamefont {Liu}, \citenamefont
  {Gray}, \citenamefont {Kareev}, \citenamefont {Moon}, \citenamefont {Prasai},
  \citenamefont {Cohn}, \citenamefont {Varela}, \citenamefont {Tung},
  \citenamefont {Bedzyk}, \citenamefont {Altendorf}, \citenamefont {Strigari},
  \citenamefont {Dabrowski}, \citenamefont {Tjeng}, \citenamefont {Ryan},\ and\
  \citenamefont {Freeland}}]{Chakhalian2011}%
  \BibitemOpen
  \bibfield  {author} {\bibinfo {author} {\bibfnamefont {J.}~\bibnamefont
  {Chakhalian}}, \bibinfo {author} {\bibfnamefont {J.~M.}\ \bibnamefont
  {Rondinelli}}, \bibinfo {author} {\bibfnamefont {J.}~\bibnamefont {Liu}},
  \bibinfo {author} {\bibfnamefont {B.~A.}\ \bibnamefont {Gray}}, \bibinfo
  {author} {\bibfnamefont {M.}~\bibnamefont {Kareev}}, \bibinfo {author}
  {\bibfnamefont {E.~J.}\ \bibnamefont {Moon}}, \bibinfo {author}
  {\bibfnamefont {N.}~\bibnamefont {Prasai}}, \bibinfo {author} {\bibfnamefont
  {J.~L.}\ \bibnamefont {Cohn}}, \bibinfo {author} {\bibfnamefont
  {M.}~\bibnamefont {Varela}}, \bibinfo {author} {\bibfnamefont {I.~C.}\
  \bibnamefont {Tung}}, \bibinfo {author} {\bibfnamefont {M.~J.}\ \bibnamefont
  {Bedzyk}}, \bibinfo {author} {\bibfnamefont {S.~G.}\ \bibnamefont
  {Altendorf}}, \bibinfo {author} {\bibfnamefont {F.}~\bibnamefont {Strigari}},
  \bibinfo {author} {\bibfnamefont {B.}~\bibnamefont {Dabrowski}}, \bibinfo
  {author} {\bibfnamefont {L.~H.}\ \bibnamefont {Tjeng}}, \bibinfo {author}
  {\bibfnamefont {P.~J.}\ \bibnamefont {Ryan}},\ and\ \bibinfo {author}
  {\bibfnamefont {J.~W.}\ \bibnamefont {Freeland}},\ }\bibfield  {title}
  {\bibinfo {title} {{Asymmetric orbital-lattice interactions in ultrathin
  correlated oxide films}},\ }\href
  {https://doi.org/10.1103/PhysRevLett.107.116805} {\bibfield  {journal}
  {\bibinfo  {journal} {Phys. Rev. Lett.}\ }\textbf {\bibinfo {volume} {107}},\
  \bibinfo {pages} {1} (\bibinfo {year} {2011})}\BibitemShut {NoStop}%
\bibitem [{\citenamefont {Disa}\ \emph {et~al.}(2015)\citenamefont {Disa},
  \citenamefont {Kumah}, \citenamefont {Malashevich}, \citenamefont {Chen},
  \citenamefont {Arena}, \citenamefont {Specht}, \citenamefont {Ismail-Beigi},
  \citenamefont {Walker},\ and\ \citenamefont {Ahn}}]{Disa2015}%
  \BibitemOpen
  \bibfield  {author} {\bibinfo {author} {\bibfnamefont {A.~S.}\ \bibnamefont
  {Disa}}, \bibinfo {author} {\bibfnamefont {D.~P.}\ \bibnamefont {Kumah}},
  \bibinfo {author} {\bibfnamefont {A.}~\bibnamefont {Malashevich}}, \bibinfo
  {author} {\bibfnamefont {H.}~\bibnamefont {Chen}}, \bibinfo {author}
  {\bibfnamefont {D.~A.}\ \bibnamefont {Arena}}, \bibinfo {author}
  {\bibfnamefont {E.~D.}\ \bibnamefont {Specht}}, \bibinfo {author}
  {\bibfnamefont {S.}~\bibnamefont {Ismail-Beigi}}, \bibinfo {author}
  {\bibfnamefont {F.~J.}\ \bibnamefont {Walker}},\ and\ \bibinfo {author}
  {\bibfnamefont {C.~H.}\ \bibnamefont {Ahn}},\ }\bibfield  {title} {\bibinfo
  {title} {{Orbital engineering in symmetry-breaking polar heterostructures}},\
  }\href {https://doi.org/10.1103/PhysRevLett.114.026801} {\bibfield  {journal}
  {\bibinfo  {journal} {Phys. Rev. Lett.}\ }\textbf {\bibinfo {volume} {114}},\
  \bibinfo {pages} {1} (\bibinfo {year} {2015})}\BibitemShut {NoStop}%
\bibitem [{\citenamefont {Benckiser}\ \emph {et~al.}(2011)\citenamefont
  {Benckiser}, \citenamefont {Haverkort}, \citenamefont {Br{\"{u}}ck},
  \citenamefont {Goering}, \citenamefont {MacKe}, \citenamefont {Fra{\~{n}}},
  \citenamefont {Yang}, \citenamefont {Andersen}, \citenamefont {Cristiani},
  \citenamefont {Habermeier}, \citenamefont {Boris}, \citenamefont
  {Zegkinoglou}, \citenamefont {Wochner}, \citenamefont {Kim}, \citenamefont
  {Hinkov},\ and\ \citenamefont {Keimer}}]{Benckiser2011}%
  \BibitemOpen
  \bibfield  {author} {\bibinfo {author} {\bibfnamefont {E.}~\bibnamefont
  {Benckiser}}, \bibinfo {author} {\bibfnamefont {M.~W.}\ \bibnamefont
  {Haverkort}}, \bibinfo {author} {\bibfnamefont {S.}~\bibnamefont
  {Br{\"{u}}ck}}, \bibinfo {author} {\bibfnamefont {E.}~\bibnamefont
  {Goering}}, \bibinfo {author} {\bibfnamefont {S.}~\bibnamefont {MacKe}},
  \bibinfo {author} {\bibfnamefont {A.}~\bibnamefont {Fra{\~{n}}}}, \bibinfo
  {author} {\bibfnamefont {X.}~\bibnamefont {Yang}}, \bibinfo {author}
  {\bibfnamefont {O.~K.}\ \bibnamefont {Andersen}}, \bibinfo {author}
  {\bibfnamefont {G.}~\bibnamefont {Cristiani}}, \bibinfo {author}
  {\bibfnamefont {H.~U.}\ \bibnamefont {Habermeier}}, \bibinfo {author}
  {\bibfnamefont {A.~V.}\ \bibnamefont {Boris}}, \bibinfo {author}
  {\bibfnamefont {I.}~\bibnamefont {Zegkinoglou}}, \bibinfo {author}
  {\bibfnamefont {P.}~\bibnamefont {Wochner}}, \bibinfo {author} {\bibfnamefont
  {H.~J.}\ \bibnamefont {Kim}}, \bibinfo {author} {\bibfnamefont
  {V.}~\bibnamefont {Hinkov}},\ and\ \bibinfo {author} {\bibfnamefont
  {B.}~\bibnamefont {Keimer}},\ }\bibfield  {title} {\bibinfo {title} {{Orbital
  reflectometry of oxide heterostructures}},\ }\href
  {https://doi.org/10.1038/nmat2958} {\bibfield  {journal} {\bibinfo  {journal}
  {Nat. Mater.}\ }\textbf {\bibinfo {volume} {10}},\ \bibinfo {pages} {189}
  (\bibinfo {year} {2011})}\BibitemShut {NoStop}%
\bibitem [{\citenamefont {Cao}\ \emph {et~al.}(2016)\citenamefont {Cao},
  \citenamefont {Liu}, \citenamefont {Kareev}, \citenamefont {Choudhury},
  \citenamefont {Middey}, \citenamefont {Meyers}, \citenamefont {Kim},
  \citenamefont {Ryan}, \citenamefont {Freeland},\ and\ \citenamefont
  {Chakhalian}}]{Cao2016}%
  \BibitemOpen
  \bibfield  {author} {\bibinfo {author} {\bibfnamefont {Y.}~\bibnamefont
  {Cao}}, \bibinfo {author} {\bibfnamefont {X.}~\bibnamefont {Liu}}, \bibinfo
  {author} {\bibfnamefont {M.}~\bibnamefont {Kareev}}, \bibinfo {author}
  {\bibfnamefont {D.}~\bibnamefont {Choudhury}}, \bibinfo {author}
  {\bibfnamefont {S.}~\bibnamefont {Middey}}, \bibinfo {author} {\bibfnamefont
  {D.}~\bibnamefont {Meyers}}, \bibinfo {author} {\bibfnamefont {J.-W.}\
  \bibnamefont {Kim}}, \bibinfo {author} {\bibfnamefont {P.~J.}\ \bibnamefont
  {Ryan}}, \bibinfo {author} {\bibfnamefont {J.~W.}\ \bibnamefont {Freeland}},\
  and\ \bibinfo {author} {\bibfnamefont {J.}~\bibnamefont {Chakhalian}},\
  }\bibfield  {title} {\bibinfo {title} {{Engineered Mott ground state in a
  LaTiO$_{3+\delta}$/LaNiO$_3$ heterostructure}},\ }\href
  {https://doi.org/10.1038/ncomms10418} {\bibfield  {journal} {\bibinfo
  {journal} {Nat. Commun.}\ }\textbf {\bibinfo {volume} {7}},\ \bibinfo {pages}
  {10418} (\bibinfo {year} {2016})}\BibitemShut {NoStop}%
\bibitem [{\citenamefont {Li}\ \emph {et~al.}(2019)\citenamefont {Li},
  \citenamefont {Lee}, \citenamefont {Wang}, \citenamefont {Osada},
  \citenamefont {Crossley}, \citenamefont {Lee}, \citenamefont {Cui},
  \citenamefont {Hikita},\ and\ \citenamefont {Hwang}}]{Li2019}%
  \BibitemOpen
  \bibfield  {author} {\bibinfo {author} {\bibfnamefont {D.}~\bibnamefont
  {Li}}, \bibinfo {author} {\bibfnamefont {K.}~\bibnamefont {Lee}}, \bibinfo
  {author} {\bibfnamefont {B.~Y.}\ \bibnamefont {Wang}}, \bibinfo {author}
  {\bibfnamefont {M.}~\bibnamefont {Osada}}, \bibinfo {author} {\bibfnamefont
  {S.}~\bibnamefont {Crossley}}, \bibinfo {author} {\bibfnamefont {H.~R.}\
  \bibnamefont {Lee}}, \bibinfo {author} {\bibfnamefont {Y.}~\bibnamefont
  {Cui}}, \bibinfo {author} {\bibfnamefont {Y.}~\bibnamefont {Hikita}},\ and\
  \bibinfo {author} {\bibfnamefont {H.~Y.}\ \bibnamefont {Hwang}},\ }\bibfield
  {title} {\bibinfo {title} {Superconductivity in an infinite-layer
  nickelate},\ }\href {https://doi.org/10.1038/s41586-019-1496-5} {\bibfield
  {journal} {\bibinfo  {journal} {Nature}\ }\textbf {\bibinfo {volume} {572}},\
  \bibinfo {pages} {624} (\bibinfo {year} {2019})}\BibitemShut {NoStop}%
\bibitem [{\citenamefont {Goodge}\ \emph {et~al.}(2021)\citenamefont {Goodge},
  \citenamefont {Li}, \citenamefont {Lee}, \citenamefont {Osada}, \citenamefont
  {Wang}, \citenamefont {Sawatzky}, \citenamefont {Hwang},\ and\ \citenamefont
  {Kourkoutis}}]{Goodgee2007683118}%
  \BibitemOpen
  \bibfield  {author} {\bibinfo {author} {\bibfnamefont {B.~H.}\ \bibnamefont
  {Goodge}}, \bibinfo {author} {\bibfnamefont {D.}~\bibnamefont {Li}}, \bibinfo
  {author} {\bibfnamefont {K.}~\bibnamefont {Lee}}, \bibinfo {author}
  {\bibfnamefont {M.}~\bibnamefont {Osada}}, \bibinfo {author} {\bibfnamefont
  {B.~Y.}\ \bibnamefont {Wang}}, \bibinfo {author} {\bibfnamefont {G.~A.}\
  \bibnamefont {Sawatzky}}, \bibinfo {author} {\bibfnamefont {H.~Y.}\
  \bibnamefont {Hwang}},\ and\ \bibinfo {author} {\bibfnamefont {L.~F.}\
  \bibnamefont {Kourkoutis}},\ }\bibfield  {title} {\bibinfo {title} {{Doping
  evolution of the Mott-Hubbard landscape in infinite-layer nickelates}},\
  }\bibfield  {journal} {\bibinfo  {journal} {Proc. Natl. Acad. Sci.}\ }\textbf
  {\bibinfo {volume} {118}},\ \href {https://doi.org/10.1073/pnas.2007683118}
  {10.1073/pnas.2007683118} (\bibinfo {year} {2021})\BibitemShut {NoStop}%
\bibitem [{\citenamefont {Hepting}\ \emph {et~al.}(2020)\citenamefont
  {Hepting}, \citenamefont {Li}, \citenamefont {Jia}, \citenamefont {Lu},
  \citenamefont {Paris}, \citenamefont {Tseng}, \citenamefont {Feng},
  \citenamefont {Osada}, \citenamefont {Been}, \citenamefont {Hikita},
  \citenamefont {Chuang}, \citenamefont {Hussain}, \citenamefont {Zhou},
  \citenamefont {Nag}, \citenamefont {Garcia-Fernandez}, \citenamefont {Rossi},
  \citenamefont {Huang}, \citenamefont {Huang}, \citenamefont {Shen},
  \citenamefont {Schmitt}, \citenamefont {Hwang}, \citenamefont {Moritz},
  \citenamefont {Zaanen}, \citenamefont {Devereaux},\ and\ \citenamefont
  {Lee}}]{Hepting2020}%
  \BibitemOpen
  \bibfield  {author} {\bibinfo {author} {\bibfnamefont {M.}~\bibnamefont
  {Hepting}}, \bibinfo {author} {\bibfnamefont {D.}~\bibnamefont {Li}},
  \bibinfo {author} {\bibfnamefont {C.~J.}\ \bibnamefont {Jia}}, \bibinfo
  {author} {\bibfnamefont {H.}~\bibnamefont {Lu}}, \bibinfo {author}
  {\bibfnamefont {E.}~\bibnamefont {Paris}}, \bibinfo {author} {\bibfnamefont
  {Y.}~\bibnamefont {Tseng}}, \bibinfo {author} {\bibfnamefont
  {X.}~\bibnamefont {Feng}}, \bibinfo {author} {\bibfnamefont {M.}~\bibnamefont
  {Osada}}, \bibinfo {author} {\bibfnamefont {E.}~\bibnamefont {Been}},
  \bibinfo {author} {\bibfnamefont {Y.}~\bibnamefont {Hikita}}, \bibinfo
  {author} {\bibfnamefont {Y.-D.}\ \bibnamefont {Chuang}}, \bibinfo {author}
  {\bibfnamefont {Z.}~\bibnamefont {Hussain}}, \bibinfo {author} {\bibfnamefont
  {K.~J.}\ \bibnamefont {Zhou}}, \bibinfo {author} {\bibfnamefont
  {A.}~\bibnamefont {Nag}}, \bibinfo {author} {\bibfnamefont {M.}~\bibnamefont
  {Garcia-Fernandez}}, \bibinfo {author} {\bibfnamefont {M.}~\bibnamefont
  {Rossi}}, \bibinfo {author} {\bibfnamefont {H.~Y.}\ \bibnamefont {Huang}},
  \bibinfo {author} {\bibfnamefont {D.~J.}\ \bibnamefont {Huang}}, \bibinfo
  {author} {\bibfnamefont {Z.~X.}\ \bibnamefont {Shen}}, \bibinfo {author}
  {\bibfnamefont {T.}~\bibnamefont {Schmitt}}, \bibinfo {author} {\bibfnamefont
  {H.~Y.}\ \bibnamefont {Hwang}}, \bibinfo {author} {\bibfnamefont
  {B.}~\bibnamefont {Moritz}}, \bibinfo {author} {\bibfnamefont
  {J.}~\bibnamefont {Zaanen}}, \bibinfo {author} {\bibfnamefont {T.~P.}\
  \bibnamefont {Devereaux}},\ and\ \bibinfo {author} {\bibfnamefont {W.~S.}\
  \bibnamefont {Lee}},\ }\bibfield  {title} {\bibinfo {title} {{Electronic
  structure of the parent compound of superconducting infinite-layer
  nickelates}},\ }\bibfield  {journal} {\bibinfo  {journal} {Nat. Mater.}\
  }\href {https://doi.org/10.1038/s41563-019-0585-z}
  {10.1038/s41563-019-0585-z} (\bibinfo {year} {2020})\BibitemShut {NoStop}%
\bibitem [{\citenamefont {Rossi}\ \emph {et~al.}(2020)\citenamefont {Rossi},
  \citenamefont {Lu}, \citenamefont {Nag}, \citenamefont {Li}, \citenamefont
  {Osada}, \citenamefont {Lee}, \citenamefont {Wang}, \citenamefont
  {Agrestini}, \citenamefont {Garcia-Fernandez}, \citenamefont {Chuang},
  \citenamefont {Shen}, \citenamefont {Hwang}, \citenamefont {Moritz},
  \citenamefont {Zhou}, \citenamefont {Devereaux},\ and\ \citenamefont
  {Lee}}]{Rossi2020}%
  \BibitemOpen
  \bibfield  {author} {\bibinfo {author} {\bibfnamefont {M.}~\bibnamefont
  {Rossi}}, \bibinfo {author} {\bibfnamefont {H.}~\bibnamefont {Lu}}, \bibinfo
  {author} {\bibfnamefont {A.}~\bibnamefont {Nag}}, \bibinfo {author}
  {\bibfnamefont {D.}~\bibnamefont {Li}}, \bibinfo {author} {\bibfnamefont
  {M.}~\bibnamefont {Osada}}, \bibinfo {author} {\bibfnamefont
  {K.}~\bibnamefont {Lee}}, \bibinfo {author} {\bibfnamefont {B.~Y.}\
  \bibnamefont {Wang}}, \bibinfo {author} {\bibfnamefont {S.}~\bibnamefont
  {Agrestini}}, \bibinfo {author} {\bibfnamefont {M.}~\bibnamefont
  {Garcia-Fernandez}}, \bibinfo {author} {\bibfnamefont {Y.~D.}\ \bibnamefont
  {Chuang}}, \bibinfo {author} {\bibfnamefont {Z.~X.}\ \bibnamefont {Shen}},
  \bibinfo {author} {\bibfnamefont {H.~Y.}\ \bibnamefont {Hwang}}, \bibinfo
  {author} {\bibfnamefont {B.}~\bibnamefont {Moritz}}, \bibinfo {author}
  {\bibfnamefont {K.-J.}\ \bibnamefont {Zhou}}, \bibinfo {author}
  {\bibfnamefont {T.~P.}\ \bibnamefont {Devereaux}},\ and\ \bibinfo {author}
  {\bibfnamefont {W.~S.}\ \bibnamefont {Lee}},\ }\bibfield  {title} {\bibinfo
  {title} {{Orbital and Spin Character of Doped Carriers in Infinite-Layer
  Nickelates}},\ }\href {http://arxiv.org/abs/2011.00595} {\bibfield  {journal}
  {\bibinfo  {journal} {arXiv}\ }\textbf {\bibinfo {volume} {2011.00595}}
  (\bibinfo {year} {2020})}\BibitemShut {NoStop}%
\bibitem [{\citenamefont {Botana}\ and\ \citenamefont
  {Norman}(2020)}]{BotanaNorman2020}%
  \BibitemOpen
  \bibfield  {author} {\bibinfo {author} {\bibfnamefont {A.~S.}\ \bibnamefont
  {Botana}}\ and\ \bibinfo {author} {\bibfnamefont {M.~R.}\ \bibnamefont
  {Norman}},\ }\bibfield  {title} {\bibinfo {title} {{Similarities and
  Differences between LaNiO$_2$ and CaCuO$_2$ and Implications for
  Superconductivity}},\ }\href {https://doi.org/10.1103/PhysRevX.10.011024}
  {\bibfield  {journal} {\bibinfo  {journal} {Phys. Rev. X}\ }\textbf {\bibinfo
  {volume} {10}},\ \bibinfo {pages} {11024} (\bibinfo {year}
  {2020})}\BibitemShut {NoStop}%
\bibitem [{\citenamefont {Zhang}\ \emph {et~al.}(2020)\citenamefont {Zhang},
  \citenamefont {Jin}, \citenamefont {Wang}, \citenamefont {Xi}, \citenamefont
  {Shi}, \citenamefont {Ye},\ and\ \citenamefont {Mei}}]{Zhang2020}%
  \BibitemOpen
  \bibfield  {author} {\bibinfo {author} {\bibfnamefont {H.}~\bibnamefont
  {Zhang}}, \bibinfo {author} {\bibfnamefont {L.}~\bibnamefont {Jin}}, \bibinfo
  {author} {\bibfnamefont {S.}~\bibnamefont {Wang}}, \bibinfo {author}
  {\bibfnamefont {B.}~\bibnamefont {Xi}}, \bibinfo {author} {\bibfnamefont
  {X.}~\bibnamefont {Shi}}, \bibinfo {author} {\bibfnamefont {F.}~\bibnamefont
  {Ye}},\ and\ \bibinfo {author} {\bibfnamefont {J.-W.}\ \bibnamefont {Mei}},\
  }\bibfield  {title} {\bibinfo {title} {{Effective Hamiltonian for nickelate
  oxides Nd$_{1-x}$Sr$_x$NiO$_2$}},\ }\href
  {https://doi.org/10.1103/PhysRevResearch.2.013214} {\bibfield  {journal}
  {\bibinfo  {journal} {Phys. Rev. Res.}\ }\textbf {\bibinfo {volume} {2}},\
  \bibinfo {pages} {13214} (\bibinfo {year} {2020})}\BibitemShut {NoStop}%
\bibitem [{\citenamefont {Karp}\ \emph {et~al.}(2020)\citenamefont {Karp},
  \citenamefont {Botana}, \citenamefont {Norman}, \citenamefont {Park},
  \citenamefont {Zingl},\ and\ \citenamefont {Millis}}]{Millis2020}%
  \BibitemOpen
  \bibfield  {author} {\bibinfo {author} {\bibfnamefont {J.}~\bibnamefont
  {Karp}}, \bibinfo {author} {\bibfnamefont {A.~S.}\ \bibnamefont {Botana}},
  \bibinfo {author} {\bibfnamefont {M.~R.}\ \bibnamefont {Norman}}, \bibinfo
  {author} {\bibfnamefont {H.}~\bibnamefont {Park}}, \bibinfo {author}
  {\bibfnamefont {M.}~\bibnamefont {Zingl}},\ and\ \bibinfo {author}
  {\bibfnamefont {A.}~\bibnamefont {Millis}},\ }\bibfield  {title} {\bibinfo
  {title} {{Many-Body Electronic Structure of NdNiO$_2$ and CaCuO$_2$}},\
  }\href {https://doi.org/10.1103/PhysRevX.10.021061} {\bibfield  {journal}
  {\bibinfo  {journal} {Phys. Rev. X}\ }\textbf {\bibinfo {volume} {10}},\
  \bibinfo {pages} {21061} (\bibinfo {year} {2020})}\BibitemShut {NoStop}%
\bibitem [{\citenamefont {Jiang}\ \emph {et~al.}(2020)\citenamefont {Jiang},
  \citenamefont {Berciu},\ and\ \citenamefont {Sawatzky}}]{Sawatzky2020}%
  \BibitemOpen
  \bibfield  {author} {\bibinfo {author} {\bibfnamefont {M.}~\bibnamefont
  {Jiang}}, \bibinfo {author} {\bibfnamefont {M.}~\bibnamefont {Berciu}},\ and\
  \bibinfo {author} {\bibfnamefont {G.~A.}\ \bibnamefont {Sawatzky}},\
  }\bibfield  {title} {\bibinfo {title} {{Critical Nature of the Ni Spin State
  in Doped NdNiO$_2$}},\ }\href
  {https://doi.org/10.1103/PhysRevLett.124.207004} {\bibfield  {journal}
  {\bibinfo  {journal} {Phys. Rev. Lett.}\ }\textbf {\bibinfo {volume} {124}},\
  \bibinfo {pages} {207004} (\bibinfo {year} {2020})}\BibitemShut {NoStop}%
\bibitem [{\citenamefont {Pellegrin}\ \emph {et~al.}(1993)\citenamefont
  {Pellegrin}, \citenamefont {N{\"{u}}cker}, \citenamefont {Fink},
  \citenamefont {Molodtsov}, \citenamefont {Guti{\'{e}}rrez}, \citenamefont
  {Navas}, \citenamefont {Strebel}, \citenamefont {Hu}, \citenamefont {Domke},
  \citenamefont {Kaindl}, \citenamefont {Uchida}, \citenamefont {Nakamura},
  \citenamefont {Markl}, \citenamefont {Klauda}, \citenamefont
  {Saemann-Ischenko}, \citenamefont {Krol}, \citenamefont {Peng}, \citenamefont
  {Li},\ and\ \citenamefont {Greene}}]{Pellegrin1993}%
  \BibitemOpen
  \bibfield  {author} {\bibinfo {author} {\bibfnamefont {E.}~\bibnamefont
  {Pellegrin}}, \bibinfo {author} {\bibfnamefont {N.}~\bibnamefont
  {N{\"{u}}cker}}, \bibinfo {author} {\bibfnamefont {J.}~\bibnamefont {Fink}},
  \bibinfo {author} {\bibfnamefont {S.~L.}\ \bibnamefont {Molodtsov}}, \bibinfo
  {author} {\bibfnamefont {A.}~\bibnamefont {Guti{\'{e}}rrez}}, \bibinfo
  {author} {\bibfnamefont {E.}~\bibnamefont {Navas}}, \bibinfo {author}
  {\bibfnamefont {O.}~\bibnamefont {Strebel}}, \bibinfo {author} {\bibfnamefont
  {Z.}~\bibnamefont {Hu}}, \bibinfo {author} {\bibfnamefont {M.}~\bibnamefont
  {Domke}}, \bibinfo {author} {\bibfnamefont {G.}~\bibnamefont {Kaindl}},
  \bibinfo {author} {\bibfnamefont {S.}~\bibnamefont {Uchida}}, \bibinfo
  {author} {\bibfnamefont {Y.}~\bibnamefont {Nakamura}}, \bibinfo {author}
  {\bibfnamefont {J.}~\bibnamefont {Markl}}, \bibinfo {author} {\bibfnamefont
  {M.}~\bibnamefont {Klauda}}, \bibinfo {author} {\bibfnamefont
  {G.}~\bibnamefont {Saemann-Ischenko}}, \bibinfo {author} {\bibfnamefont
  {A.}~\bibnamefont {Krol}}, \bibinfo {author} {\bibfnamefont {J.~L.}\
  \bibnamefont {Peng}}, \bibinfo {author} {\bibfnamefont {Z.~Y.}\ \bibnamefont
  {Li}},\ and\ \bibinfo {author} {\bibfnamefont {R.~L.}\ \bibnamefont
  {Greene}},\ }\bibfield  {title} {\bibinfo {title} {{Orbital character of
  states at the Fermi level in La$_{2-x}$Sr$_x$CuO$_4$ and
  R$_{2-x}$Ce$_x$CuO$_4$}},\ }\href {https://doi.org/10.1103/PhysRevB.47.3354}
  {\bibfield  {journal} {\bibinfo  {journal} {Phys. Rev. B}\ }\textbf {\bibinfo
  {volume} {47}},\ \bibinfo {pages} {3354} (\bibinfo {year}
  {1993})}\BibitemShut {NoStop}%
\bibitem [{\citenamefont {Brookes}\ \emph {et~al.}(2001)\citenamefont
  {Brookes}, \citenamefont {Ghiringhelli}, \citenamefont {Tjernberg},
  \citenamefont {Tjeng}, \citenamefont {Mizokawa}, \citenamefont {Li},\ and\
  \citenamefont {Menovsky}}]{Brookes2001}%
  \BibitemOpen
  \bibfield  {author} {\bibinfo {author} {\bibfnamefont {N.~B.}\ \bibnamefont
  {Brookes}}, \bibinfo {author} {\bibfnamefont {G.}~\bibnamefont
  {Ghiringhelli}}, \bibinfo {author} {\bibfnamefont {O.}~\bibnamefont
  {Tjernberg}}, \bibinfo {author} {\bibfnamefont {L.~H.}\ \bibnamefont
  {Tjeng}}, \bibinfo {author} {\bibfnamefont {T.}~\bibnamefont {Mizokawa}},
  \bibinfo {author} {\bibfnamefont {T.~W.}\ \bibnamefont {Li}},\ and\ \bibinfo
  {author} {\bibfnamefont {A.~A.}\ \bibnamefont {Menovsky}},\ }\bibfield
  {title} {\bibinfo {title} {{Detection of Zhang-Rice Singlets Using
  Spin-Polarized Photoemission}},\ }\href
  {https://doi.org/10.1103/PhysRevLett.87.237003} {\bibfield  {journal}
  {\bibinfo  {journal} {Phys. Rev. Lett.}\ }\textbf {\bibinfo {volume} {87}},\
  \bibinfo {pages} {237003} (\bibinfo {year} {2001})}\BibitemShut {NoStop}%
\bibitem [{\citenamefont {Lu}\ \emph {et~al.}(2021)\citenamefont {Lu},
  \citenamefont {Rossi}, \citenamefont {Nag}, \citenamefont {Osada},
  \citenamefont {Li}, \citenamefont {Lee}, \citenamefont {Wang}, \citenamefont
  {Garcia-Fernandez}, \citenamefont {Agrestini}, \citenamefont {Shen},
  \citenamefont {Been}, \citenamefont {Moritz}, \citenamefont {Devereaux},
  \citenamefont {Zaanen}, \citenamefont {Hwang}, \citenamefont {Zhou},\ and\
  \citenamefont {Lee}}]{Lu213}%
  \BibitemOpen
  \bibfield  {author} {\bibinfo {author} {\bibfnamefont {H.}~\bibnamefont
  {Lu}}, \bibinfo {author} {\bibfnamefont {M.}~\bibnamefont {Rossi}}, \bibinfo
  {author} {\bibfnamefont {A.}~\bibnamefont {Nag}}, \bibinfo {author}
  {\bibfnamefont {M.}~\bibnamefont {Osada}}, \bibinfo {author} {\bibfnamefont
  {D.~F.}\ \bibnamefont {Li}}, \bibinfo {author} {\bibfnamefont
  {K.}~\bibnamefont {Lee}}, \bibinfo {author} {\bibfnamefont {B.~Y.}\
  \bibnamefont {Wang}}, \bibinfo {author} {\bibfnamefont {M.}~\bibnamefont
  {Garcia-Fernandez}}, \bibinfo {author} {\bibfnamefont {S.}~\bibnamefont
  {Agrestini}}, \bibinfo {author} {\bibfnamefont {Z.~X.}\ \bibnamefont {Shen}},
  \bibinfo {author} {\bibfnamefont {E.~M.}\ \bibnamefont {Been}}, \bibinfo
  {author} {\bibfnamefont {B.}~\bibnamefont {Moritz}}, \bibinfo {author}
  {\bibfnamefont {T.~P.}\ \bibnamefont {Devereaux}}, \bibinfo {author}
  {\bibfnamefont {J.}~\bibnamefont {Zaanen}}, \bibinfo {author} {\bibfnamefont
  {H.~Y.}\ \bibnamefont {Hwang}}, \bibinfo {author} {\bibfnamefont {K.-J.}\
  \bibnamefont {Zhou}},\ and\ \bibinfo {author} {\bibfnamefont {W.~S.}\
  \bibnamefont {Lee}},\ }\bibfield  {title} {\bibinfo {title} {{Magnetic
  excitations in infinite-layer nickelates}},\ }\href
  {https://doi.org/10.1126/science.abd7726} {\bibfield  {journal} {\bibinfo
  {journal} {Science (80).}\ }\textbf {\bibinfo {volume} {373}},\ \bibinfo
  {pages} {213} (\bibinfo {year} {2021})}\BibitemShut {NoStop}%
\bibitem [{\citenamefont {Zeng}\ \emph {et~al.}(2020)\citenamefont {Zeng},
  \citenamefont {Tang}, \citenamefont {Yin}, \citenamefont {Li}, \citenamefont
  {Li}, \citenamefont {Huang}, \citenamefont {Hu}, \citenamefont {Liu},
  \citenamefont {Omar}, \citenamefont {Jani}, \citenamefont {Lim},
  \citenamefont {Han}, \citenamefont {Wan}, \citenamefont {Yang}, \citenamefont
  {Pennycook}, \citenamefont {Wee},\ and\ \citenamefont
  {Ariando}}]{Ariando2020}%
  \BibitemOpen
  \bibfield  {author} {\bibinfo {author} {\bibfnamefont {S.}~\bibnamefont
  {Zeng}}, \bibinfo {author} {\bibfnamefont {C.~S.}\ \bibnamefont {Tang}},
  \bibinfo {author} {\bibfnamefont {X.}~\bibnamefont {Yin}}, \bibinfo {author}
  {\bibfnamefont {C.}~\bibnamefont {Li}}, \bibinfo {author} {\bibfnamefont
  {M.}~\bibnamefont {Li}}, \bibinfo {author} {\bibfnamefont {Z.}~\bibnamefont
  {Huang}}, \bibinfo {author} {\bibfnamefont {J.}~\bibnamefont {Hu}}, \bibinfo
  {author} {\bibfnamefont {W.}~\bibnamefont {Liu}}, \bibinfo {author}
  {\bibfnamefont {G.~J.}\ \bibnamefont {Omar}}, \bibinfo {author}
  {\bibfnamefont {H.}~\bibnamefont {Jani}}, \bibinfo {author} {\bibfnamefont
  {Z.~S.}\ \bibnamefont {Lim}}, \bibinfo {author} {\bibfnamefont
  {K.}~\bibnamefont {Han}}, \bibinfo {author} {\bibfnamefont {D.}~\bibnamefont
  {Wan}}, \bibinfo {author} {\bibfnamefont {P.}~\bibnamefont {Yang}}, \bibinfo
  {author} {\bibfnamefont {S.~J.}\ \bibnamefont {Pennycook}}, \bibinfo {author}
  {\bibfnamefont {A.~T.~S.}\ \bibnamefont {Wee}},\ and\ \bibinfo {author}
  {\bibfnamefont {A.}~\bibnamefont {Ariando}},\ }\bibfield  {title} {\bibinfo
  {title} {{Phase Diagram and Superconducting Dome of Infinite-Layer
  Nd$_{1-x}$Sr$_x$NiO$_2$ Thin Films}},\ }\href
  {https://doi.org/10.1103/PhysRevLett.125.147003} {\bibfield  {journal}
  {\bibinfo  {journal} {Phys. Rev. Lett.}\ }\textbf {\bibinfo {volume} {125}},\
  \bibinfo {pages} {147003} (\bibinfo {year} {2020})}\BibitemShut {NoStop}%
\bibitem [{\citenamefont {Emery}\ and\ \citenamefont
  {Kivelson}(1993)}]{emery1993frustrated}%
  \BibitemOpen
  \bibfield  {author} {\bibinfo {author} {\bibfnamefont {V.~J.}\ \bibnamefont
  {Emery}}\ and\ \bibinfo {author} {\bibfnamefont {S.}~\bibnamefont
  {Kivelson}},\ }\bibfield  {title} {\bibinfo {title} {Frustrated electronic
  phase separation and high-temperature superconductors},\ }\href@noop {}
  {\bibfield  {journal} {\bibinfo  {journal} {Physica C}\ }\textbf {\bibinfo
  {volume} {209}},\ \bibinfo {pages} {597} (\bibinfo {year}
  {1993})}\BibitemShut {NoStop}%
\bibitem [{\citenamefont {Castellani}\ \emph {et~al.}(1995)\citenamefont
  {Castellani}, \citenamefont {Di~Castro},\ and\ \citenamefont
  {Grilli}}]{castellani1995singular}%
  \BibitemOpen
  \bibfield  {author} {\bibinfo {author} {\bibfnamefont {C.}~\bibnamefont
  {Castellani}}, \bibinfo {author} {\bibfnamefont {C.}~\bibnamefont
  {Di~Castro}},\ and\ \bibinfo {author} {\bibfnamefont {M.}~\bibnamefont
  {Grilli}},\ }\bibfield  {title} {\bibinfo {title} {Singular quasiparticle
  scattering in the proximity of charge instabilities},\ }\href@noop {}
  {\bibfield  {journal} {\bibinfo  {journal} {Phys. Rev. Lett.}\ }\textbf
  {\bibinfo {volume} {75}},\ \bibinfo {pages} {4650} (\bibinfo {year}
  {1995})}\BibitemShut {NoStop}%
\bibitem [{\citenamefont {Castellani}\ \emph {et~al.}(1996)\citenamefont
  {Castellani}, \citenamefont {Di~Castro},\ and\ \citenamefont
  {Grilli}}]{castellani1996non}%
  \BibitemOpen
  \bibfield  {author} {\bibinfo {author} {\bibfnamefont {C.}~\bibnamefont
  {Castellani}}, \bibinfo {author} {\bibfnamefont {C.}~\bibnamefont
  {Di~Castro}},\ and\ \bibinfo {author} {\bibfnamefont {M.}~\bibnamefont
  {Grilli}},\ }\bibfield  {title} {\bibinfo {title} {Non-fermi-liquid behavior
  and d-wave superconductivity near the charge-density-wave quantum critical
  point},\ }\href@noop {} {\bibfield  {journal} {\bibinfo  {journal} {Z. Phys.
  B}\ }\textbf {\bibinfo {volume} {103}},\ \bibinfo {pages} {137} (\bibinfo
  {year} {1996})}\BibitemShut {NoStop}%
\bibitem [{\citenamefont {Ghiringhelli}\ \emph {et~al.}(2012)\citenamefont
  {Ghiringhelli}, \citenamefont {{Le Tacon}}, \citenamefont {Minola},
  \citenamefont {Blanco-Canosa}, \citenamefont {Mazzoli}, \citenamefont
  {Brookes}, \citenamefont {{De Luca}}, \citenamefont {Frano}, \citenamefont
  {Hawthorn}, \citenamefont {He}, \citenamefont {Loew}, \citenamefont {Sala},
  \citenamefont {Peets}, \citenamefont {Salluzzo}, \citenamefont {Schierle},
  \citenamefont {Sutarto}, \citenamefont {Sawatzky}, \citenamefont {Weschke},
  \citenamefont {Keimer},\ and\ \citenamefont {Braicovich}}]{Ghiringhelli821}%
  \BibitemOpen
  \bibfield  {author} {\bibinfo {author} {\bibfnamefont {G.}~\bibnamefont
  {Ghiringhelli}}, \bibinfo {author} {\bibfnamefont {M.}~\bibnamefont {{Le
  Tacon}}}, \bibinfo {author} {\bibfnamefont {M.}~\bibnamefont {Minola}},
  \bibinfo {author} {\bibfnamefont {S.}~\bibnamefont {Blanco-Canosa}}, \bibinfo
  {author} {\bibfnamefont {C.}~\bibnamefont {Mazzoli}}, \bibinfo {author}
  {\bibfnamefont {N.~B.}\ \bibnamefont {Brookes}}, \bibinfo {author}
  {\bibfnamefont {G.~M.}\ \bibnamefont {{De Luca}}}, \bibinfo {author}
  {\bibfnamefont {A.}~\bibnamefont {Frano}}, \bibinfo {author} {\bibfnamefont
  {D.~G.}\ \bibnamefont {Hawthorn}}, \bibinfo {author} {\bibfnamefont
  {F.}~\bibnamefont {He}}, \bibinfo {author} {\bibfnamefont {T.}~\bibnamefont
  {Loew}}, \bibinfo {author} {\bibfnamefont {M.~M.}\ \bibnamefont {Sala}},
  \bibinfo {author} {\bibfnamefont {D.~C.}\ \bibnamefont {Peets}}, \bibinfo
  {author} {\bibfnamefont {M.}~\bibnamefont {Salluzzo}}, \bibinfo {author}
  {\bibfnamefont {E.}~\bibnamefont {Schierle}}, \bibinfo {author}
  {\bibfnamefont {R.}~\bibnamefont {Sutarto}}, \bibinfo {author} {\bibfnamefont
  {G.~A.}\ \bibnamefont {Sawatzky}}, \bibinfo {author} {\bibfnamefont
  {E.}~\bibnamefont {Weschke}}, \bibinfo {author} {\bibfnamefont
  {B.}~\bibnamefont {Keimer}},\ and\ \bibinfo {author} {\bibfnamefont
  {L.}~\bibnamefont {Braicovich}},\ }\bibfield  {title} {\bibinfo {title}
  {{Long-Range Incommensurate Charge Fluctuations in (Y,Nd)Ba2Cu3O6+x}},\
  }\href {https://doi.org/10.1126/science.1223532} {\bibfield  {journal}
  {\bibinfo  {journal} {Science}\ }\textbf {\bibinfo {volume} {337}},\ \bibinfo
  {pages} {821} (\bibinfo {year} {2012})}\BibitemShut {NoStop}%
\bibitem [{\citenamefont {Achkar}\ \emph {et~al.}(2012)\citenamefont {Achkar},
  \citenamefont {Sutarto}, \citenamefont {Mao}, \citenamefont {He},
  \citenamefont {Frano}, \citenamefont {Blanco-Canosa}, \citenamefont
  {Le~Tacon}, \citenamefont {Ghiringhelli}, \citenamefont {Braicovich},
  \citenamefont {Minola} \emph {et~al.}}]{achkar2012distinct}%
  \BibitemOpen
  \bibfield  {author} {\bibinfo {author} {\bibfnamefont {A.}~\bibnamefont
  {Achkar}}, \bibinfo {author} {\bibfnamefont {R.}~\bibnamefont {Sutarto}},
  \bibinfo {author} {\bibfnamefont {X.}~\bibnamefont {Mao}}, \bibinfo {author}
  {\bibfnamefont {F.}~\bibnamefont {He}}, \bibinfo {author} {\bibfnamefont
  {A.}~\bibnamefont {Frano}}, \bibinfo {author} {\bibfnamefont
  {S.}~\bibnamefont {Blanco-Canosa}}, \bibinfo {author} {\bibfnamefont
  {M.}~\bibnamefont {Le~Tacon}}, \bibinfo {author} {\bibfnamefont
  {G.}~\bibnamefont {Ghiringhelli}}, \bibinfo {author} {\bibfnamefont
  {L.}~\bibnamefont {Braicovich}}, \bibinfo {author} {\bibfnamefont
  {M.}~\bibnamefont {Minola}}, \emph {et~al.},\ }\bibfield  {title} {\bibinfo
  {title} {Distinct charge orders in the planes and chains of ortho-iii-ordered
  yba 2 cu 3 o 6+ $\delta$ superconductors identified by resonant elastic x-ray
  scattering},\ }\href@noop {} {\bibfield  {journal} {\bibinfo  {journal}
  {Phys. Rev. Lett.}\ }\textbf {\bibinfo {volume} {109}},\ \bibinfo {pages}
  {167001} (\bibinfo {year} {2012})}\BibitemShut {NoStop}%
\bibitem [{\citenamefont {Comin}\ and\ \citenamefont
  {Damascelli}(2016)}]{CD2016}%
  \BibitemOpen
  \bibfield  {author} {\bibinfo {author} {\bibfnamefont {R.}~\bibnamefont
  {Comin}}\ and\ \bibinfo {author} {\bibfnamefont {A.}~\bibnamefont
  {Damascelli}},\ }\bibfield  {title} {\bibinfo {title} {{Resonant X-Ray
  Scattering Studies of Charge Order in Cuprates}},\ }\href
  {https://doi.org/10.1146/annurev-conmatphys-031115-011401} {\bibfield
  {journal} {\bibinfo  {journal} {Annu. Rev. Condens. Matter Phys.}\ }\textbf
  {\bibinfo {volume} {7}},\ \bibinfo {pages} {369} (\bibinfo {year}
  {2016})}\BibitemShut {NoStop}%
\bibitem [{\citenamefont {Wahlberg}\ \emph {et~al.}(2021)\citenamefont
  {Wahlberg}, \citenamefont {Arpaia}, \citenamefont {Seibold}, \citenamefont
  {Rossi}, \citenamefont {Fumagalli}, \citenamefont {Trabaldo}, \citenamefont
  {Brookes}, \citenamefont {Braicovich}, \citenamefont {Caprara}, \citenamefont
  {Gran} \emph {et~al.}}]{wahlberg2021restored}%
  \BibitemOpen
  \bibfield  {author} {\bibinfo {author} {\bibfnamefont {E.}~\bibnamefont
  {Wahlberg}}, \bibinfo {author} {\bibfnamefont {R.}~\bibnamefont {Arpaia}},
  \bibinfo {author} {\bibfnamefont {G.}~\bibnamefont {Seibold}}, \bibinfo
  {author} {\bibfnamefont {M.}~\bibnamefont {Rossi}}, \bibinfo {author}
  {\bibfnamefont {R.}~\bibnamefont {Fumagalli}}, \bibinfo {author}
  {\bibfnamefont {E.}~\bibnamefont {Trabaldo}}, \bibinfo {author}
  {\bibfnamefont {N.~B.}\ \bibnamefont {Brookes}}, \bibinfo {author}
  {\bibfnamefont {L.}~\bibnamefont {Braicovich}}, \bibinfo {author}
  {\bibfnamefont {S.}~\bibnamefont {Caprara}}, \bibinfo {author} {\bibfnamefont
  {U.}~\bibnamefont {Gran}}, \emph {et~al.},\ }\bibfield  {title} {\bibinfo
  {title} {Restored strange metal phase through suppression of charge density
  waves in underdoped yba$_2$cu$_3$o$_{7-\delta}$},\ }\href
  {https://doi.org/10.1126/science.abc8372} {\bibfield  {journal} {\bibinfo
  {journal} {Science}\ }\textbf {\bibinfo {volume} {373}},\ \bibinfo {pages}
  {1506} (\bibinfo {year} {2021})}\BibitemShut {NoStop}%
\bibitem [{\citenamefont {Arpaia}\ \emph {et~al.}(2019)\citenamefont {Arpaia},
  \citenamefont {Caprara}, \citenamefont {Fumagalli}, \citenamefont {{De
  Vecchi}}, \citenamefont {Peng}, \citenamefont {Andersson}, \citenamefont
  {Betto}, \citenamefont {{De Luca}}, \citenamefont {Brookes}, \citenamefont
  {Lombardi}, \citenamefont {Salluzzo}, \citenamefont {Braicovich},
  \citenamefont {{Di Castro}}, \citenamefont {Grilli},\ and\ \citenamefont
  {Ghiringhelli}}]{Arpaia906}%
  \BibitemOpen
  \bibfield  {author} {\bibinfo {author} {\bibfnamefont {R.}~\bibnamefont
  {Arpaia}}, \bibinfo {author} {\bibfnamefont {S.}~\bibnamefont {Caprara}},
  \bibinfo {author} {\bibfnamefont {R.}~\bibnamefont {Fumagalli}}, \bibinfo
  {author} {\bibfnamefont {G.}~\bibnamefont {{De Vecchi}}}, \bibinfo {author}
  {\bibfnamefont {Y.~Y.}\ \bibnamefont {Peng}}, \bibinfo {author}
  {\bibfnamefont {E.}~\bibnamefont {Andersson}}, \bibinfo {author}
  {\bibfnamefont {D.}~\bibnamefont {Betto}}, \bibinfo {author} {\bibfnamefont
  {G.~M.}\ \bibnamefont {{De Luca}}}, \bibinfo {author} {\bibfnamefont {N.~B.}\
  \bibnamefont {Brookes}}, \bibinfo {author} {\bibfnamefont {F.}~\bibnamefont
  {Lombardi}}, \bibinfo {author} {\bibfnamefont {M.}~\bibnamefont {Salluzzo}},
  \bibinfo {author} {\bibfnamefont {L.}~\bibnamefont {Braicovich}}, \bibinfo
  {author} {\bibfnamefont {C.}~\bibnamefont {{Di Castro}}}, \bibinfo {author}
  {\bibfnamefont {M.}~\bibnamefont {Grilli}},\ and\ \bibinfo {author}
  {\bibfnamefont {G.}~\bibnamefont {Ghiringhelli}},\ }\bibfield  {title}
  {\bibinfo {title} {{Dynamical charge density fluctuations pervading the phase
  diagram of a Cu-based high-Tc superconductor}},\ }\href
  {https://doi.org/10.1126/science.aav1315} {\bibfield  {journal} {\bibinfo
  {journal} {Science}\ }\textbf {\bibinfo {volume} {365}},\ \bibinfo {pages}
  {906} (\bibinfo {year} {2019})}\BibitemShut {NoStop}%
\bibitem [{\citenamefont {Seibold}\ \emph {et~al.}(2021)\citenamefont
  {Seibold}, \citenamefont {Arpaia}, \citenamefont {Peng}, \citenamefont
  {Fumagalli}, \citenamefont {Braicovich}, \citenamefont {Di~Castro},
  \citenamefont {Grilli}, \citenamefont {Ghiringhelli},\ and\ \citenamefont
  {Caprara}}]{seibold2021strange}%
  \BibitemOpen
  \bibfield  {author} {\bibinfo {author} {\bibfnamefont {G.}~\bibnamefont
  {Seibold}}, \bibinfo {author} {\bibfnamefont {R.}~\bibnamefont {Arpaia}},
  \bibinfo {author} {\bibfnamefont {Y.~Y.}\ \bibnamefont {Peng}}, \bibinfo
  {author} {\bibfnamefont {R.}~\bibnamefont {Fumagalli}}, \bibinfo {author}
  {\bibfnamefont {L.}~\bibnamefont {Braicovich}}, \bibinfo {author}
  {\bibfnamefont {C.}~\bibnamefont {Di~Castro}}, \bibinfo {author}
  {\bibfnamefont {M.}~\bibnamefont {Grilli}}, \bibinfo {author} {\bibfnamefont
  {G.~C.}\ \bibnamefont {Ghiringhelli}},\ and\ \bibinfo {author} {\bibfnamefont
  {S.}~\bibnamefont {Caprara}},\ }\bibfield  {title} {\bibinfo {title} {Strange
  metal behaviour from charge density fluctuations in cuprates},\ }\href@noop
  {} {\bibfield  {journal} {\bibinfo  {journal} {Commun. Phys.}\ }\textbf
  {\bibinfo {volume} {4}},\ \bibinfo {pages} {7} (\bibinfo {year}
  {2021})}\BibitemShut {NoStop}%
\bibitem [{\citenamefont {Arpaia}\ and\ \citenamefont
  {Ghiringhelli}(2021)}]{Riccardo}%
  \BibitemOpen
  \bibfield  {author} {\bibinfo {author} {\bibfnamefont {R.}~\bibnamefont
  {Arpaia}}\ and\ \bibinfo {author} {\bibfnamefont {G.}~\bibnamefont
  {Ghiringhelli}},\ }\bibfield  {title} {\bibinfo {title} {{Charge Order at
  High Temperature in Cuprate Superconductors}},\ }\href
  {https://doi.org/10.7566/JPSJ.90.111005} {\bibfield  {journal} {\bibinfo
  {journal} {J. Phys. Soc. Japan}\ }\textbf {\bibinfo {volume} {90}},\ \bibinfo
  {pages} {111005} (\bibinfo {year} {2021})}\BibitemShut {NoStop}%
\bibitem [{\citenamefont {Preziosi}\ \emph {et~al.}(2017)\citenamefont
  {Preziosi}, \citenamefont {Sander}, \citenamefont {Barth{\'{e}}l{\'{e}}my},\
  and\ \citenamefont {Bibes}}]{Preziosi2017}%
  \BibitemOpen
  \bibfield  {author} {\bibinfo {author} {\bibfnamefont {D.}~\bibnamefont
  {Preziosi}}, \bibinfo {author} {\bibfnamefont {A.}~\bibnamefont {Sander}},
  \bibinfo {author} {\bibfnamefont {A.}~\bibnamefont
  {Barth{\'{e}}l{\'{e}}my}},\ and\ \bibinfo {author} {\bibfnamefont
  {M.}~\bibnamefont {Bibes}},\ }\bibfield  {title} {\bibinfo {title}
  {{Reproducibility and off-stoichiometry issues in nickelate thin films grown
  by pulsed laser deposition}},\ }\href {https://doi.org/10.1063/1.4975307}
  {\bibfield  {journal} {\bibinfo  {journal} {AIP Adv.}\ }\textbf {\bibinfo
  {volume} {7}},\ \bibinfo {pages} {015210} (\bibinfo {year}
  {2017})}\BibitemShut {NoStop}%
\bibitem [{SI()}]{SI}%
  \BibitemOpen
  \href {...} {\emph {\bibinfo {title} {{See Supplemental Material for further
  experimental details at}}}}\BibitemShut {NoStop}%
\bibitem [{\citenamefont {Li}\ \emph {et~al.}(2020)\citenamefont {Li},
  \citenamefont {He}, \citenamefont {Si}, \citenamefont {Zhu}, \citenamefont
  {Zhang},\ and\ \citenamefont {Wen}}]{Li2020}%
  \BibitemOpen
  \bibfield  {author} {\bibinfo {author} {\bibfnamefont {Q.}~\bibnamefont
  {Li}}, \bibinfo {author} {\bibfnamefont {C.}~\bibnamefont {He}}, \bibinfo
  {author} {\bibfnamefont {J.}~\bibnamefont {Si}}, \bibinfo {author}
  {\bibfnamefont {X.}~\bibnamefont {Zhu}}, \bibinfo {author} {\bibfnamefont
  {Y.}~\bibnamefont {Zhang}},\ and\ \bibinfo {author} {\bibfnamefont {H.-H.}\
  \bibnamefont {Wen}},\ }\bibfield  {title} {\bibinfo {title} {{Absence of
  superconductivity in bulk Nd$_{1−x}$Sr$_x$NiO$_2$}},\ }\href
  {https://doi.org/10.1038/s43246-020-0018-1} {\bibfield  {journal} {\bibinfo
  {journal} {Commun. Mater.}\ }\textbf {\bibinfo {volume} {1}},\ \bibinfo
  {pages} {16} (\bibinfo {year} {2020})}\BibitemShut {NoStop}%
\bibitem [{\citenamefont {Brookes}\ \emph {et~al.}(2018)\citenamefont
  {Brookes}, \citenamefont {Yakhou-Harris}, \citenamefont {Kummer},
  \citenamefont {Fondacaro}, \citenamefont {Cezar}, \citenamefont {Betto},
  \citenamefont {Velez-Fort}, \citenamefont {Amorese}, \citenamefont
  {Ghiringhelli}, \citenamefont {Braicovich}, \citenamefont {Barrett},
  \citenamefont {Berruyer}, \citenamefont {Cianciosi}, \citenamefont {Eybert},
  \citenamefont {Marion}, \citenamefont {van~der Linden},\ and\ \citenamefont
  {Zhang}}]{Brookes2018}%
  \BibitemOpen
  \bibfield  {author} {\bibinfo {author} {\bibfnamefont {N.~B.}\ \bibnamefont
  {Brookes}}, \bibinfo {author} {\bibfnamefont {F.}~\bibnamefont
  {Yakhou-Harris}}, \bibinfo {author} {\bibfnamefont {K.}~\bibnamefont
  {Kummer}}, \bibinfo {author} {\bibfnamefont {A.}~\bibnamefont {Fondacaro}},
  \bibinfo {author} {\bibfnamefont {J.~C.}\ \bibnamefont {Cezar}}, \bibinfo
  {author} {\bibfnamefont {D.}~\bibnamefont {Betto}}, \bibinfo {author}
  {\bibfnamefont {E.}~\bibnamefont {Velez-Fort}}, \bibinfo {author}
  {\bibfnamefont {A.}~\bibnamefont {Amorese}}, \bibinfo {author} {\bibfnamefont
  {G.}~\bibnamefont {Ghiringhelli}}, \bibinfo {author} {\bibfnamefont
  {L.}~\bibnamefont {Braicovich}}, \bibinfo {author} {\bibfnamefont
  {R.}~\bibnamefont {Barrett}}, \bibinfo {author} {\bibfnamefont
  {G.}~\bibnamefont {Berruyer}}, \bibinfo {author} {\bibfnamefont
  {F.}~\bibnamefont {Cianciosi}}, \bibinfo {author} {\bibfnamefont
  {L.}~\bibnamefont {Eybert}}, \bibinfo {author} {\bibfnamefont
  {P.}~\bibnamefont {Marion}}, \bibinfo {author} {\bibfnamefont
  {P.}~\bibnamefont {van~der Linden}},\ and\ \bibinfo {author} {\bibfnamefont
  {L.}~\bibnamefont {Zhang}},\ }\bibfield  {title} {\bibinfo {title} {{The
  beamline ID32 at the ESRF for soft X-ray high energy resolution resonant
  inelastic X-ray scattering and polarisation dependent X-ray absorption
  spectroscopy}},\ }\href
  {https://doi.org/https://doi.org/10.1016/j.nima.2018.07.001} {\bibfield
  {journal} {\bibinfo  {journal} {Nucl. Instrum. Methods A}\ }\textbf {\bibinfo
  {volume} {903}},\ \bibinfo {pages} {175} (\bibinfo {year}
  {2018})}\BibitemShut {NoStop}%
\bibitem [{\citenamefont {Peng}\ \emph {et~al.}(2018)\citenamefont {Peng},
  \citenamefont {Fumagalli}, \citenamefont {Ding}, \citenamefont {Minola},
  \citenamefont {Caprara}, \citenamefont {Betto}, \citenamefont {Bluschke},
  \citenamefont {{De Luca}}, \citenamefont {Kummer}, \citenamefont
  {Lefran{\c{c}}ois}, \citenamefont {Salluzzo}, \citenamefont {Suzuki},
  \citenamefont {{Le Tacon}}, \citenamefont {Zhou}, \citenamefont {Brookes},
  \citenamefont {Keimer}, \citenamefont {Braicovich}, \citenamefont {Grilli},\
  and\ \citenamefont {Ghiringhelli}}]{Peng2018}%
  \BibitemOpen
  \bibfield  {author} {\bibinfo {author} {\bibfnamefont {Y.~Y.}\ \bibnamefont
  {Peng}}, \bibinfo {author} {\bibfnamefont {R.}~\bibnamefont {Fumagalli}},
  \bibinfo {author} {\bibfnamefont {Y.}~\bibnamefont {Ding}}, \bibinfo {author}
  {\bibfnamefont {M.}~\bibnamefont {Minola}}, \bibinfo {author} {\bibfnamefont
  {S.}~\bibnamefont {Caprara}}, \bibinfo {author} {\bibfnamefont
  {D.}~\bibnamefont {Betto}}, \bibinfo {author} {\bibfnamefont
  {M.}~\bibnamefont {Bluschke}}, \bibinfo {author} {\bibfnamefont {G.~M.}\
  \bibnamefont {{De Luca}}}, \bibinfo {author} {\bibfnamefont {K.}~\bibnamefont
  {Kummer}}, \bibinfo {author} {\bibfnamefont {E.}~\bibnamefont
  {Lefran{\c{c}}ois}}, \bibinfo {author} {\bibfnamefont {M.}~\bibnamefont
  {Salluzzo}}, \bibinfo {author} {\bibfnamefont {H.}~\bibnamefont {Suzuki}},
  \bibinfo {author} {\bibfnamefont {M.}~\bibnamefont {{Le Tacon}}}, \bibinfo
  {author} {\bibfnamefont {X.~J.}\ \bibnamefont {Zhou}}, \bibinfo {author}
  {\bibfnamefont {N.~B.}\ \bibnamefont {Brookes}}, \bibinfo {author}
  {\bibfnamefont {B.}~\bibnamefont {Keimer}}, \bibinfo {author} {\bibfnamefont
  {L.}~\bibnamefont {Braicovich}}, \bibinfo {author} {\bibfnamefont
  {M.}~\bibnamefont {Grilli}},\ and\ \bibinfo {author} {\bibfnamefont
  {G.}~\bibnamefont {Ghiringhelli}},\ }\bibfield  {title} {\bibinfo {title}
  {{Re-entrant charge order in overdoped
  (Bi,Pb)$_{2.12}$Sr$_{1.88}$CuO$_{6+\delta}$ outside the pseudogap regime}},\
  }\href {https://doi.org/10.1038/s41563-018-0108-3} {\bibfield  {journal}
  {\bibinfo  {journal} {Nat. Mater.}\ }\textbf {\bibinfo {volume} {17}},\
  \bibinfo {pages} {697} (\bibinfo {year} {2018})}\BibitemShut {NoStop}%
\bibitem [{\citenamefont {Si}\ \emph {et~al.}(2020)\citenamefont {Si},
  \citenamefont {Xiao}, \citenamefont {Kaufmann}, \citenamefont {Tomczak},
  \citenamefont {Lu}, \citenamefont {Zhong},\ and\ \citenamefont
  {Held}}]{TopoTheory2020}%
  \BibitemOpen
  \bibfield  {author} {\bibinfo {author} {\bibfnamefont {L.}~\bibnamefont
  {Si}}, \bibinfo {author} {\bibfnamefont {W.}~\bibnamefont {Xiao}}, \bibinfo
  {author} {\bibfnamefont {J.}~\bibnamefont {Kaufmann}}, \bibinfo {author}
  {\bibfnamefont {J.~M.}\ \bibnamefont {Tomczak}}, \bibinfo {author}
  {\bibfnamefont {Y.}~\bibnamefont {Lu}}, \bibinfo {author} {\bibfnamefont
  {Z.}~\bibnamefont {Zhong}},\ and\ \bibinfo {author} {\bibfnamefont
  {K.}~\bibnamefont {Held}},\ }\bibfield  {title} {\bibinfo {title}
  {{Topotactic Hydrogen in Nickelate Superconductors and Akin Infinite-Layer
  Oxides ABO$_2$}},\ }\href {https://doi.org/10.1103/PhysRevLett.124.166402}
  {\bibfield  {journal} {\bibinfo  {journal} {Phys. Rev. Lett.}\ }\textbf
  {\bibinfo {volume} {124}},\ \bibinfo {pages} {166402} (\bibinfo {year}
  {2020})}\BibitemShut {NoStop}%
\end{thebibliography}%
\end{document}